\documentclass[prd,twocolumn,twoside,preprintnumbers,superscriptaddress,nofootinbib]{revtex4}
\usepackage{amsmath,slashed}
\usepackage{graphicx,graphics,color}
\usepackage{dcolumn}
\usepackage[hyperfootnotes=false]{hyperref}
\usepackage{xspace}

\newcommand{\GeV}{\,\text{GeV}}

\newcommand{\diff}{\text{d}}

\newcommand{\beq}{\begin{equation}}
\newcommand{\eeq}{\end{equation}}
\renewcommand{\Re}{\text{Re}\,}

\def\Xint#1{\mathchoice
   {\XXint\displaystyle\textstyle{#1}}%
   {\XXint\textstyle\scriptstyle{#1}}%
   {\XXint\scriptstyle\scriptscriptstyle{#1}}%
   {\XXint\scriptscriptstyle\scriptscriptstyle{#1}}%
   \!\int}
\def\XXint#1#2#3{{\setbox0=\hbox{$#1{#2#3}{\int}$}
     \vcenter{\hbox{$#2#3$}}\kern-0.5\wd0}}
\providecommand{\dashint}[1][0pt]{\Xint{\hspace{#1}-}}

\allowdisplaybreaks[1]

\begin{document}

\preprint{PSI-PR-20-04, UZ-TH 06/20}
\title{Hadronic vacuum polarization: $\boldsymbol{(g-2)_\mu}$ versus global electroweak fits}

\author{Andreas Crivellin}
\email{andreas.crivellin@cern.ch}
\affiliation{Paul Scherrer Institut, CH--5232 Villigen PSI, Switzerland}
\affiliation{Physik-Institut, Universit\"at Z\"urich, Winterthurerstrasse 190, CH--8057 Z\"urich, Switzerland}
\author{Martin Hoferichter}
\email{hoferichter@itp.unibe.ch}
\affiliation{Albert Einstein Center for Fundamental Physics, Institute for Theoretical Physics, University of Bern, Sidlerstrasse 5, CH--3012 Bern, Switzerland}

\author{Claudio Andrea Manzari}
\email{claudioandrea.manzari@physik.uzh.ch}
\affiliation{Paul Scherrer Institut, CH--5232 Villigen PSI, Switzerland}
\affiliation{Physik-Institut, Universit\"at Z\"urich, Winterthurerstrasse 190, CH--8057 Z\"urich, Switzerland}

\author{Marc Montull}
\email{marc.montull@gmail.com}
\affiliation{Paul Scherrer Institut, CH--5232 Villigen PSI, Switzerland}
\affiliation{Physik-Institut, Universit\"at Z\"urich, Winterthurerstrasse 190, CH--8057 Z\"urich, Switzerland}

\begin{abstract}
Hadronic vacuum polarization (HVP) is not only a critical part of the Standard Model (SM) prediction for the anomalous magnetic moment of the muon $(g-2)_\mu$, but also a crucial ingredient for global fits to electroweak (EW) precision observables due to its contribution to the running of the fine-structure constant encoded in $\Delta\alpha^{(5)}_\text{had}$. 
We find that with modern EW precision data, including the measurement of the Higgs mass, the global fit alone provides a competitive, independent determination of $\Delta \alpha^{(5)}_\text{had}\big|_\text{EW}=270.2(3.0)\times 10^{-4}$. This value actually lies below the range derived from $e^+e^-\to\text{hadrons}$ cross-section data, and thus goes into the opposite direction as would be required if a change in HVP were to bring the SM prediction for $(g-2)_\mu$ into agreement with the Brookhaven measurement.
Depending on the energy where the bulk of the changes in the cross section occurs, reconciling experiment and SM prediction for $(g-2)_\mu$ by adjusting HVP would thus not necessarily weaken the case for physics beyond the SM (BSM), but to some extent shift it from $(g-2)_\mu$ to the EW fit.  We briefly explore some options of BSM scenarios that could conceivably explain the ensuing tension.
\end{abstract}

\maketitle

{\it Introduction}.---The SM of particle physics has been established with increasing precision over the last decades. In particular, both the global fits to EW precision data~\cite{deBlas:2016ojx,Haller:2018nnx,Aaltonen:2018dxj} and to the Cabibbo--Kobayashi--Maskawa (CKM) matrix~\cite{Ciuchini:2000de,Hocker:2001xe} are in general in good agreement with the SM hypothesis and no new particles have been directly observed so far at the large hadron collider (LHC)~\cite{Butler:2017afk,Masetti:2018btj}.

However, low-energy precision experiments have accumulated intriguing hints for the violation of lepton flavor universality within recent years (see, e.g., Refs.~\cite{Amhis:2019ckw,Crivellin:2018gzw,Murgui:2019czp,Shi:2019gxi,Blanke:2019qrx,Kumbhakar:2019avh,Alguero:2019ptt,Aebischer:2019mlg,Ciuchini:2019usw,Coutinho:2019aiy,Crivellin:2020lzu} for $b\to c\tau\nu$, $b\to s\ell^+\ell^-$, and $R(V_{us})$). In particular, the Brookhaven measurement of the anomalous magnetic moment of the muon $(g-2)_\mu$~\cite{Bennett:2006fi}
shows a tension of about $3.7\,\sigma$ with the SM prediction~\cite{Aoyama:2020ynm,Aoyama:2012wk,Aoyama:2019ryr,Czarnecki:2002nt,Gnendiger:2013pva,Davier:2017zfy,Keshavarzi:2018mgv,Colangelo:2018mtw,Hoferichter:2019gzf,Davier:2019can,Keshavarzi:2019abf,Kurz:2014wya,Melnikov:2003xd,Masjuan:2017tvw,Colangelo:2017qdm,Colangelo:2017fiz,Hoferichter:2018dmo,Hoferichter:2018kwz,Gerardin:2019vio,Bijnens:2019ghy,Colangelo:2019lpu,Colangelo:2019uex,Blum:2019ugy,Colangelo:2014qya}.\footnote{For the electron an analogous $2.5\,\sigma$ difference (but with opposite sign) between the SM prediction~\cite{Laporta:2017okg,Aoyama:2017uqe} based on the Cs measurement of the fine-structure constant $\alpha$~\cite{Parker:2018vye} and the direct measurement of $(g-2)_e$~\cite{Hanneke:2008tm} has emerged~\cite{Davoudiasl:2018fbb,Crivellin:2018qmi}.}
Here, the QED~\cite{Aoyama:2012wk,Aoyama:2019ryr} and EW~\cite{Czarnecki:2002nt,Gnendiger:2013pva} contributions are well under control, so that the accuracy that can be achieved in testing the SM rests on the hadronic contributions. Traditionally, HVP has been determined via a dispersion relation from the cross section $\sigma(e^+e^-\to\text{hadrons})$~\cite{Bouchiat:1961lbg,Brodsky:1967sr} 
\begin{align}
\label{amu_HVP}
 a_\mu^\text{HVP}&=\bigg(\frac{\alpha m_\mu}{3\pi}\bigg)^2\int_{s_\text{thr}}^\infty \diff s \frac{\hat K(s)}{s^2}R_\text{had}(s),\notag\\
 R_\text{had}(s)&=\frac{3s}{4\pi\alpha^2}\sigma(e^+e^-\to\text{hadrons}),
\end{align}
where in the usual conventions for isospin-breaking effects the integral starts at the threshold $s_\text{thr}=M_{\pi^0}^2$ due to the $e^+e^-\to\pi^0\gamma$ channel~\cite{Hoid:2020xjs} and  the kernel function $\hat K(s)$ can be expressed analytically. Global analyses based on a direct integration of cross-section data~\cite{Davier:2017zfy,Keshavarzi:2018mgv,Davier:2019can,Keshavarzi:2019abf} can now also be combined with analyticity and unitarity constraints for the leading $2\pi$~\cite{Davier:2019can,Colangelo:2018mtw,Ananthanarayan:2018nyx} and $3\pi$~\cite{Hoferichter:2019gzf} channels, covering almost $80\%$ of the HVP contribution, to demonstrate that the experimental data sets are consistent with general properties of QCD, and radiative corrections for the $2\pi$ channel have been completed at next-to-leading order~\cite{Campanario:2019mjh}. With recent advances in constraining the contribution from hadronic light-by-light scattering (including evaluations~\cite{Colangelo:2014dfa,Colangelo:2014pva,Colangelo:2015ama,Masjuan:2017tvw,Colangelo:2017qdm,Colangelo:2017fiz,Hoferichter:2018dmo,Hoferichter:2018kwz} based on dispersion relations in analogy to Eq.~\eqref{amu_HVP}, short-distance constraints~\cite{Bijnens:2019ghy,Colangelo:2019lpu,Colangelo:2019uex}, and lattice QCD~\cite{Gerardin:2019vio,Blum:2019ugy}) as well as higher-order hadronic corrections~\cite{Calmet:1976kd,Keshavarzi:2019abf,Kurz:2014wya,Colangelo:2014qya}, this data-driven determination of HVP has corroborated the $(g-2)_\mu$ tension at the level of $3.7\,\sigma$.

Nevertheless, since by far the largest hadronic correction arises from HVP, requirements for the relative precision are extraordinary, with $a_\mu^\text{HVP}=693.1(4.0)\times 10^{-10}$~\cite{Aoyama:2020ynm,Davier:2017zfy,Keshavarzi:2018mgv,Colangelo:2018mtw,Hoferichter:2019gzf,Davier:2019can,Keshavarzi:2019abf} as currently determined from $e^+e^-\to\text{hadrons}$ cross sections corresponding to less than $0.6\%$. One may thus ask what would happen if the SM prediction were brought into agreement with experiment by changing $a_\mu^\text{HVP}$. As first discussed in Ref.~\cite{Passera:2008jk}, there is a correlation with the hadronic contribution to the running of the fine-structure constant, whose extent depends on the energy range where most of the changes occur. Here, we study this interplay
in light of modern EW precision data, including the Higgs mass, and work out the consequences for the EW fit. 

\begin{figure}[t]
	\centering
	\includegraphics[width=\linewidth]{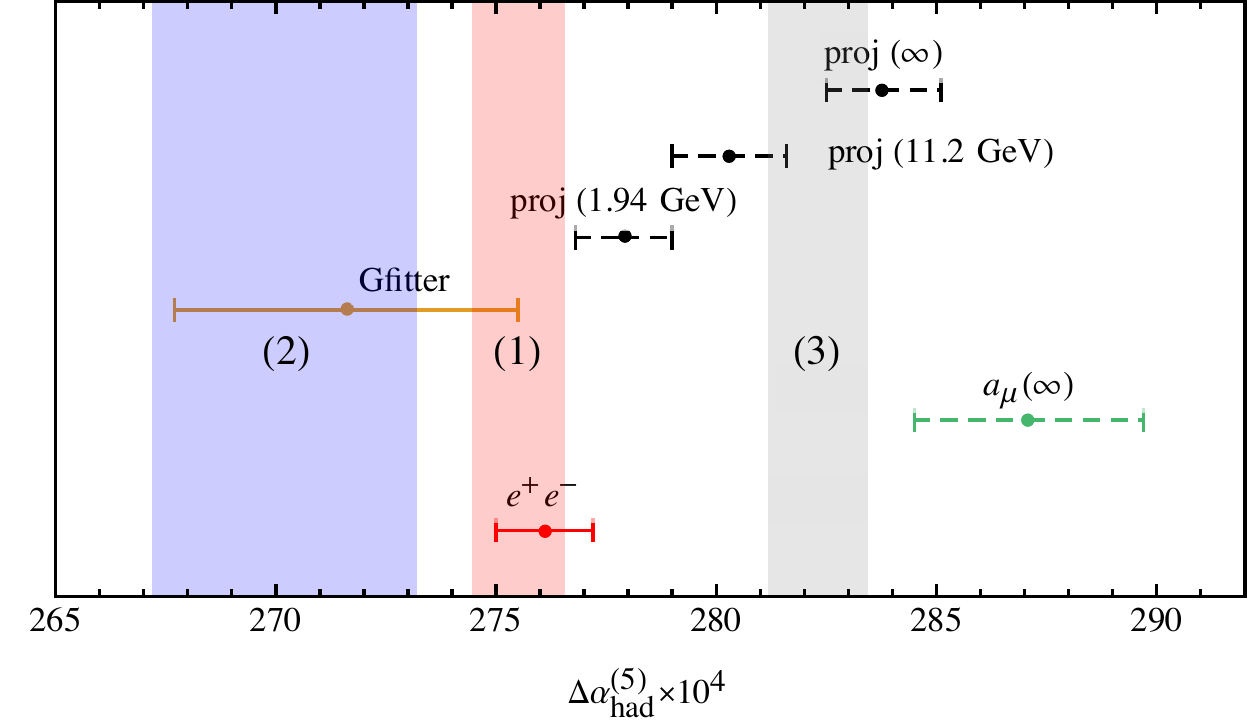}
	\caption{Summary of the different determinations of $\Delta \alpha^{(5)}_\text{had}$ ($1\,\sigma$). Results that assume the relative change in the cross section to be energy independent (compared to the $e^+e^-$ data and below the  scale indicated in brackets, as explained below Eqs.~\eqref{BMWc}--\eqref{BMWclow}) are shown as dashed lines. The colored bands indicate the posteriors within scenario (1), (2), and (3), corresponding to using $e^+e^-$ data, no input for the prior, and employing the projection~\eqref{BMWc}, respectively. In addition, we show the 2018 result for the EW fit by the Gfitter group~\cite{Haller:2018nnx}, which agrees well with our posterior (2), see Eq.~\eqref{EW}, but would slightly reduce the significance of the tension. The value derived from $a_\mu=(g-2)_\mu/2$ is obtained when assuming the absence of BSM physics in $a_\mu$ altogether and relies on the same scaling assumption as Eq.~\eqref{BMWc}, see Eq.~\eqref{g_2}.}
	\label{Delta_alpha_had}
\end{figure}

HVP enters the global EW fit indirectly via its impact on the running of $\alpha$. With $\alpha$ most accurately determined as $\alpha\equiv\alpha(0)$, but EW precision data taken around the $Z$ pole, the translation
\begin{align}
 \alpha^{-1}(M_Z^2)&=\alpha^{-1}\Big[1-\Delta\alpha_\text{lep}(M_Z^2)\notag\\
 &\qquad-\Delta \alpha^{(5)}_\text{had}(M_Z^2)-\Delta \alpha_\text{top}(M_Z^2)\Big]
 \end{align}
requires, in addition to the leptonic running $\Delta\alpha_\text{lep}$, a contribution from the top quark $\Delta \alpha_\text{top}$ and, crucially, information on the hadronic running
\begin{equation}
\label{had_running}
 \Delta \alpha^{(5)}_\text{had}(M_Z^2)=\frac{\alpha M_Z^2}{3\pi}\dashint[0.5pt]^\infty_{s_\text{thr}}\diff s \frac{R_\text{had}(s)}{s(M_Z^2-s)},
\end{equation}
where the dash indicates the principal value of the integral. Apart from a different weight function, this quantity is therefore determined by the same $e^+e^-$ cross sections, leading to the reference value~\cite{Davier:2019can,Keshavarzi:2019abf} 
\beq
\label{epem}
\Delta \alpha^{(5)}_\text{had}\big|_{e^+e^-}=276.1(1.1)\times 10^{-4}.
\eeq
In this Letter, we will study a scenario in which HVP is changed in such a way that the SM prediction for $a_\mu$ agrees with experiment within $1\,\sigma$, at a similar level of precision as currently obtained from $e^+e^-$ data, if the rest of the SM prediction remains as in Ref.~\cite{Aoyama:2020ynm}.\footnote{For definiteness, we take $a_\mu^\text{HVP}=712.5(4.5)\times 10^{-10}$, but the conclusions apply to any scenario along these lines.} Moreover, we will consider three different projections
\begin{align}
\label{BMWc}
\Delta \alpha^{(5)}_\text{had}\big|_\text{proj, $\infty$}&=283.8(1.3)\times 10^{-4},\\
\Delta \alpha^{(5)}_\text{had}\big|_\text{proj, $\leq 11.2\GeV$}&=280.3(1.3)\times 10^{-4},\label{BMWcmid}\\
\Delta \alpha^{(5)}_\text{had}\big|_\text{proj, $\leq 1.94\GeV$}&=277.9(1.1)\times 10^{-4},\label{BMWclow}
\end{align}
see Fig.~\ref{Delta_alpha_had},
which are obtained under the hypothesis 
that the relative change in the cross section occurs only below the indicated scale, but is otherwise energy independent. They thus correspond to three qualitatively different cases: Eq.~\eqref{BMWclow}, where the changes are concentrated at low energies for which HVP is determined as the sum of exclusive channels; Eq.~\eqref{BMWcmid}, where the changes extend up to energies still subject to nonperturbative contributions; and Eq.~\eqref{BMWc}, where the change would affect all energies, including those where the contribution is expected to be well described by perturbation theory.  
For definiteness, the projections~\eqref{BMWc}--\eqref{BMWclow} have been derived using the integral breakdown from Ref.~\cite{Keshavarzi:2018mgv} (Ref.~\cite{Davier:2019can} would lead to the same qualitative conclusion, but considers slightly different energy intervals). 
The significance of the tension with Eq.~\eqref{epem} becomes $\{4.5,2.5,4.5\}\sigma$ for the three cases, respectively, where in the last case the significance increases again because the dominant uncertainty in the $e^+e^-$ cross sections arising from the intermediate energy interval drops out (the remaining uncertainty is only $0.3\times 10^{-4}$~\cite{Keshavarzi:2018mgv}).
To illustrate the maximum impact on the EW fit, we will use the projection 
in Eq.~\eqref{BMWc} as a reference point, keeping mind that it should be considered an upper limit given that the perturbative contributions are unlikely to be altered.

To assess the consequences of the assumed shift in HVP, we now contrast $\Delta \alpha^{(5)}_\text{had}$ from Eqs.~\eqref{epem} and~\eqref{BMWc} to a global fit of EW precision data. We find that with modern data and theory calculations the EW fit is sufficiently powerful to provide an independent determination of $\Delta \alpha^{(5)}_\text{had}$, without assuming any prior input. We will perform this determination using the Bayesian statistics implemented in the \texttt{HEPfit} package~\cite{deBlas:2019okz}. 

{\it Electroweak Fit and HVP}.---Measurements of the EW observables, as performed at LEP~\cite{Schael:2013ita,ALEPH:2005ab}, are high-precision tests of the SM. The EW sector of the SM can be completely parameterized in terms of the three Lagrangian parameters $v$, $g$, and $g'$; then, other quantities such as the Fermi constant $G_F$ and the gauge-boson masses $M_W$, $M_Z$ can be expressed in terms of these parameters and their measurements allow for global consistency tests. However, for practical purposes it is more advantageous to choose instead the three quantities with the smallest (relative) experimental error of their direct measurements, i.e., the mass of the $Z$ boson ($M_Z$), the Fermi constant ($G_F$), and the fine-structure constant ($\alpha$). Other EW observables, computed from $G_F$, $M_Z$, and $\alpha$, include $M_W$, the hadronic $Z$-pole cross section ($\sigma_h^{0}$), and the leptonic vector and axial-vector couplings, $g_V^\ell$ and $g_A^\ell$. Assuming the gauge sector to be lepton flavor universal we can thus use the five standard $Z$ observables~\cite{ALEPH:2005ab}: $M_Z$, $\Gamma_Z$, $\sigma_{\rm had}^0$, $R_\ell^0$, and $A_{\rm FB}^{0,\ell}$. Furthermore, the Higgs mass ($M_H$), the top mass ($m_t$), and the strong coupling constant ($\alpha_s$) have to be included as fit parameters as well, since they enter indirectly EW observables via loop effects.

Similarly, $\Delta \alpha^{(5)}_\text{had}$ enters indirectly to encode the hadronic information needed to evolve $\alpha(\mu^2)$ from $\mu=0$, where its most precise measurements are performed, to the scale $\mu=M_Z$, where it is needed for the EW fit. A key new development compared to Ref.~\cite{Passera:2008jk} is that with modern EW input, especially a definite Higgs mass $M_H$, 
the EW fit is now sufficiently over-constrained that it is possible to actually determine $\Delta \alpha^{(5)}_\text{had}$ from the fit~\cite{Haller:2018nnx}. 
 Furthermore, using $\Delta \alpha^{(5)}_\text{had}$ from $e^+ e^-$ data or from our projections as an input, one can compare the goodness of the resulting fit and analyze the tensions (pulls) within the fit. We consider three different scenarios: (1) EW fit using $\Delta \alpha^{(5)}_\text{had}\big|_{e^+e^-}$ from $e^+e^-$ data as a prior; (2) EW fit without any experimental or theoretical constraint on $\Delta \alpha^{(5)}_\text{had}$ (using a large flat prior), with the posterior of $\Delta \alpha^{(5)}_\text{had}\big|_\text{EW}$ solely (albeit indirectly) determined by EW precision data;
(3) EW fit with the most extreme projection~\eqref{BMWc} as a prior for $\Delta \alpha^{(5)}_\text{had}$. Note that scenario (1) corresponds to the standard approach used previously in the literature. 

We perform the global fit within these three scenarios in a Bayesian framework using the publicly available \texttt{HEPfit} package~\cite{deBlas:2019okz}, whose Markov Chain Monte Carlo (MCMC) determination of posteriors is powered by the Bayesian Analysis Toolkit (\texttt{BAT})~\cite{Caldwell:2008fw}. The results of the three scenarios are shown in Table~\ref{Fit}, see~\cite{SuppTable}. In scenario (1) we find consistency between the value from $e^+e^-$ data and the other observables of the global fit, as can be seen from the good agreement between the measurement and the posterior of $\Delta \alpha^{(5)}_\text{had}$. In scenario (2) we find a posterior of
\begin{equation}
\label{EW}
\Delta \alpha^{(5)}_\text{had}\big|_\text{EW}=270.2(3.0)\times 10^{-4}.
\end{equation}
Note that this value (see Fig.~\ref{Delta_alpha_had} for the comparison with other determinations) has a larger error than the one obtained in scenario (1) because no additional input has been used and its posterior is entirely determined (indirectly) from the global EW fit. Our value is compatible with the 2018 Gfitter result of $271.6(3.9)\times 10^{-4}$~\cite{Haller:2018nnx}. In particular, we observe that this independent determination~\eqref{EW} of the hadronic running largely agrees with Eq.~\eqref{epem}, but differs from Eq.~\eqref{BMWc} at the level of $4.2\,\sigma$, demonstrating that if the changes to the cross section were equally distributed over all energies,
reconciling theory and experiment for $(g-2)_\mu$ in this manner
would stand in significant conflict with the EW fit. In contrast, the projections~\eqref{BMWcmid} and \eqref{BMWclow} would imply a tension of $3.1\,\sigma$ and $2.4\,\sigma$, respectively, to be compared to the $e^+e^-$ result~\eqref{epem} at $1.8\,\sigma$ above Eq.~\eqref{EW}. The same conclusion also derives from scenario (3), in which posterior and measurement of $\Delta \alpha^{(5)}_\text{had}$ are no longer in good agreement. Furthermore, the pulls of several measurements are significantly increased compared to scenario (1), signaling significant tensions within the EW fit. These tensions within scenario (3) are also confirmed by its information criterion (IC) value~\cite{Jeffreys:1998,Kass:1995} of 36, which is significantly higher than the IC values of scenarios (1) and (2) of 20.5 and 17, respectively. In the terms defined in Ref.~\cite{Kass:1995}, this constitutes ``very strong'' evidence for scenarios~(1) and (2) compared to scenario~(3).

\begin{table*}[t!]
		\begin{tabular}{l c r | r r | r r |r r} 
			\cline{4-9}
			 &  &  & \multicolumn{2}{c}{IC = 20.5}  & \multicolumn{2}{c}{IC = 17.8}  & \multicolumn{2}{c}{IC = 36.7} \\
			\toprule
			Observable & Reference & Measurement &Posterior (1)  & Pull (1) & Posterior (2) & Pull (2) & Posterior (3) & Pull (3) \\
			\colrule
			$\alpha_s(M_Z)$ & ~\cite{Tanabashi:2018oca} & $0.1181(11)$ & $0.1181(10)$ & $0.003$ & $0.1181(10)$ & $0.004$ & $0.1181(10)$& $0.02$\\
			$M_Z\,[\text{GeV}]$ &~\cite{ALEPH:2005ab}& $91.1875(21)$ & $91.1883(20)$ & $-0.27$ & $91.1877(21)$& $-0.05$ & $91.1891(20)$ & $-0.55$ \\
			$m_t\,[\text{GeV}]$ & ~\cite{TevatronElectroweakWorkingGroup:2016lid,Aaboud:2018zbu,Sirunyan:2018mlv}& $172.80(40)$ & $172.95(39)$ & $-0.27$ & $172.85(39)$ &$-0.09$ & $173.09(39)$ & $0.51$\\
			$M_H\,[\text{GeV}]$ &~\cite{Aaboud:2018wps,CMS:2019drq}  & $125.16(13)$ & $125.16(13)$ & $0.01$ & $125.16(13)$ & $0.01$ & $125.16(13)$ & $0.02$\\
			\colrule
			$M_W\,[\text{GeV}]$ & ~\cite{Tanabashi:2018oca} & $80.379(12)$ & $80.363(4)$ &$1.25$ & $80.372(6)$ & $0.56$ & $80.353(4)$ & $2.10$\\
			$\Gamma_W\,[\text{GeV}]$ & ~\cite{Tanabashi:2018oca} & $2.085(42)$ & $2.088(1)$ & $-0.09$ & $2.089(1)$ & $-0.10$ & $2.088(1)$ & $-0.07$\\
			$\text{BR}(W\to\ell\nu)$ & ~\cite{Tanabashi:2018oca} & $0.1086(9)$ & $0.10838(2)$ & $0.25$ & $0.10838(1)$& $0.25$ & $0.10838(1)$ & $0.25$ \\
			$\text{BR}(W\to \text{had})$ & ~\cite{Tanabashi:2018oca} & $0.6741(27)$ & $0.6749(1)$ & $-0.28$ & $0.6749(1)$ & $-0.28$ & $0.6749(1)$ & $-0.28$\\
			$\text{sin}^2\theta_{\rm eff}^{\rm lept}(Q^{\rm had}_{\rm FB})$& ~\cite{ALEPH:2005ab}  & $0.2324(12)$ & $0.2316(4)$ & $0.63$ & $0.2315(1)$  & $0.77$ & $0.2319(1)$ & 0.44 \\
			$\text{sin}^2\theta_{\rm eff(Had.coll.)}^{\rm lept}$ & ~\cite{Aaltonen:2018dxj,deBlas:2019okz}  & $0.23143(27)$ & $0.2316(4)$ & $-0.78$ & $0.2315(1)$  & $-0.14$  & $0.2319(1)$ & $-1.62$\\
			$P_{\tau}^{\rm pol}$ &~\cite{ALEPH:2005ab} &$0.1465(33)$ & $0.1461(3)$& $0.13$ & $0.1475(8)$& $-0.28$ & $0.1443(3)$& $0.68$\\
			$A_{\ell}$ &~\cite{ALEPH:2005ab} &$0.1513(21)$ & $0.1461(3)$ & $2.47$ & $0.1475(8)$& $1.71$ & $0.1443(3)$ & 3.31\\
			$\Gamma_Z\,[\text{GeV}]$ &~\cite{ALEPH:2005ab} &$2.4952(23)$ & $2.4947(6)$ &$0.22$ & $2.4951(6)$ & $0.05$ & $2.4942(6)$& $0.43$ \\
			$\sigma_h^{0}\,[\text{nb}]$ &~\cite{ALEPH:2005ab} &$41.541(37)$ & $41.485(6)$ & $1.50$ & $41.485(6)$& $1.51$ & $41.485(6)$ & $1.50$\\
			$R^0_{\ell}$ &~\cite{ALEPH:2005ab} &$20.767(35)$ & $20.747(7)$ & $0.79$ & $20.750(7)$ & $0.66$ & $20.743(7)$ & $0.95$\\
			$A_{\rm FB}^{0,\ell}$&~\cite{ALEPH:2005ab} &$0.0171(10)$ & $0.0160(1)$ & $1.10$ & $0.0163(2)$ & $0.78$ & $0.0156(1)$& $1.49$\\
			$R_{b}^{0}$ &~\cite{ALEPH:2005ab} &$0.21629(66)$ & $0.21582(1)$ & 0.71& $0.21582(1)$ & $0.71$ & $0.21583(1)$ & $0.70$\\
			$R_{c}^{0}$ &~\cite{ALEPH:2005ab} &$0.1721(30)$ & $0.17219(2)$& $-0.03$ & $0.17220(2)$ & $-0.03$ & $0.17218(2)$ & $-0.03$\\
			$A_{\rm FB}^{0,b}$ &~\cite{ALEPH:2005ab} &$0.0992(16)$ & $0.1024(2)$ & $-1.97$ & $0.1034(6)$ & $-2.46$ & $0.1011(2)$ & $-1.17$\\
			$A_{\rm FB}^{0,c}$ &~\cite{ALEPH:2005ab} &$0.0707(35)$ & $0.0731(2)$ & $-0.69$ & $0.0739(4)$ & $-0.90$ & $0.0721(2)$ & $-0.41$\\
			$A_{b}$ &~\cite{ALEPH:2005ab} &$0.923(20)$ & $0.93456(3)$ & $-0.58$ & $0.9347(1)$ & $-0.58$ & $0.93442(3)$ & $-0.57$\\
			$A_{c}$ &~\cite{ALEPH:2005ab} &$0.670(27)$ & $0.6675(1)$ & $0.09$ & $0.6681(4)$ & $0.07$ & $0.6667(2)$& $0.12$\\
			\botrule
	\end{tabular}
	\caption{In addition to the values given in the table we used 
		$G_F\,[\text{GeV}^{-2}]=1.1663787\times 10^{-5}$~\cite{Tanabashi:2018oca,Tishchenko:2012ie} and
		$\alpha=7.2973525698\times 10^{-3}$~\cite{Tanabashi:2018oca}, which are so precisely measured that the posteriors are identical to their direct measurements. Concerning the  $W$ mass computation, \texttt{HEPfit} provides both the option of using the precise numerical formula from Ref.~\cite{Awramik:2003rn} as well as the usual determination of $M_W$ from  $G_F$, $M_Z$, and $\alpha$~\cite{Sirlin:1980nh}, with radiative corrections encoded in $\Delta r$ (which is known up to 3-loop $\mathcal{O}(\alpha^3)$ EW~\cite{Faisst:2003px} and $\mathcal{O}(\alpha \alpha_s^2,\, \alpha^2 \alpha_s)$ EW--QCD contributions~\cite{Faisst:2003px,Avdeev:1994db,Chetyrkin:1995ix,Chetyrkin:1995js}). We opt for the latter possibility.
		\label{Fit}}
\end{table*}

{\it BSM physics in the EW fit}.---As demonstrated most conclusively in terms of Eq.~\eqref{EW}, removing the tension between SM prediction and experiment for $(g-2)_\mu$ 
by a change in HVP increases the existing tensions within the EW fit. Thus, the hints for BSM physics are difficult to be removed in this way, but always shifted at least to some extent from $(g-2)_\mu$ to the EW fit. Therefore, the question arises if there are BSM scenarios that would impact the EW fit in the observed manner, while leaving $(g-2)_\mu$ unaffected.  

As can be seen from Table~\ref{Fit}, the main tensions (largest pulls) of the fit in scenario (3) are in the $W$ mass and even more pronounced in
\begin{equation}
A_\ell =\frac{2\,\Re\big[g_V^\ell/g_A^\ell\big]}{1+\left(\Re\big[g_V^\ell/g_A^\ell\big]\right)^2},
\end{equation}
where $g_A^\ell$ ($g_V^\ell$) is the axial-vector (vector) coupling of charged leptons to the $Z$~\cite{Tanabashi:2018oca}. Another notable pull in scenario (3) appears in $\text{sin}^2\theta_{\rm eff(Had.coll.)}^{\rm lept}$, while the pull in $A_{\rm FB}^{0,b}$, the second-most significant one in the standard fit, is one of the few that becomes mitigated.

In order to get a shift in $A_\ell$, an effect in $g_V^\ell/g_A^\ell$ is necessary. In the EFT language~\cite{Buchmuller:1985jz,Grzadkowski:2010es}, this shift can be generated by effects from the operators $O_{\phi e}^{fi}$, $O_{\phi \ell}^{(1)fi}$, and $O_{\phi \ell}^{(3)fi}$. At tree level, these operators can be modified by vector-like leptons or a $Z^\prime$ boson coupling to right-handed leptons and mixing with the SM $Z$~\cite{Langacker:1991pg,deBlas:2017xtg}. Furthermore, these effects are expected to affect the closely related observable $A^{0,\ell}_{\rm FB}$ as well, where also a tension in scenario (3) arises.				
					
Concerning the $W$ mass, this shift can be understood as an effect in the EW $T$ parameter~\cite{Kennedy:1988sn,Altarelli:1990zd,Grinstein:1991cd,Peskin:1991sw} generated by $O_{\phi D}$. Here, a possible explanation could be given in terms of the minimal supersymmetric SM (MSSM), where a necessarily constructive effect (increasing the value of $M_W$ with respect to the SM) is predicted~\cite{Heinemeyer:2013dia} as confirmed by current fits~\cite{Bagnaschi:2017tru}. Furthermore, composite Higgs models have been known for a long time to be prime candidates to solve the EW hierarchy problem, and can give rise to sizable effects in the EW precision data, in particular in the $S$ and $T$ parameters~\cite{Holdom:1990tc,Golden:1990ig, Giudice:2007fh, Panico:2015jxa}. Usually, to protect tree-level modifications of the $T$ parameter, custodial symmetry is imposed. Nonetheless, its value can still be substantially modified via fermion resonances, as shown for instance in Refs.~\cite{Barbieri:2012tu,Giudice:2007fh, Panico:2015jxa}.

One could go even further and determine HVP by demanding exact agreement (within the uncertainties) between experiment and the remaining part of the SM prediction. This means that $(g-2)_\mu$ measurements could be used to determine HVP under the assumption that it is free of BSM effects and, more crucially, assuming a certain energy dependence of the changes in the cross section. 
A naive scaling with respect to Eq.~\eqref{epem} would lead to  
\beq
\label{g_2}
\Delta \alpha^{(5)}_\text{had}\big|_\text{$(g-2)_\mu$, $\infty$}=287.1(2.6) \times 10^{-4},
\eeq
by definition even larger than Eq.~\eqref{BMWc}, and with an error that would decrease to about $1.0$ for the final E989 precision~\cite{Grange:2015fou}. 
The comparison of the different values for $\Delta \alpha^{(5)}_\text{had}$ is shown in Fig.~\ref{Delta_alpha_had}, with the ones affected by the scaling assumption indicated by dashed lines. In view of these different scenarios it is worthwhile to assess the impact of future determinations of $\Delta \alpha^{(5)}_\text{had}$ on the global EW fit. 

\begin{figure}[t]
	\centering
\hspace{0cm}\includegraphics[scale=0.451]{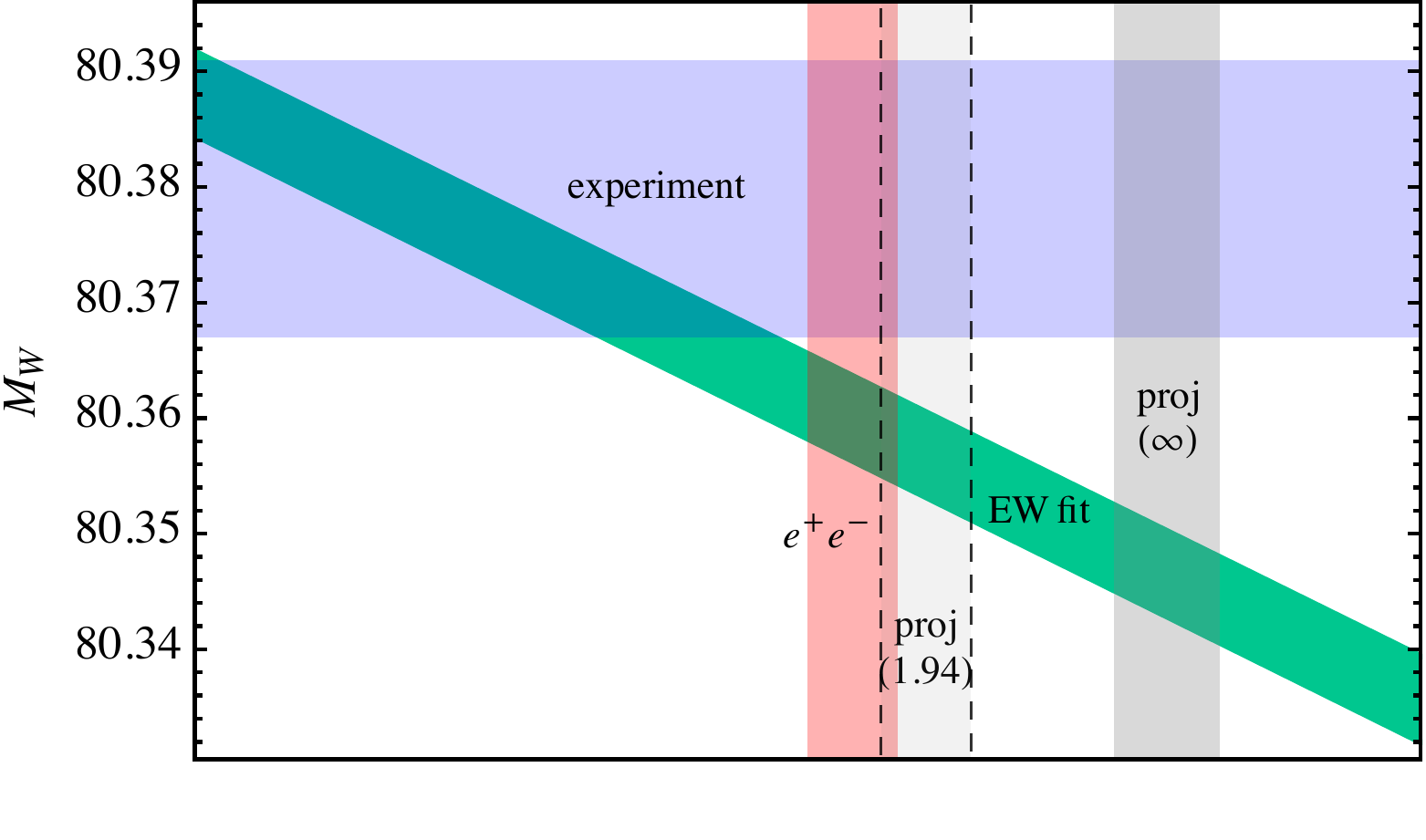}\\[-0.4cm]
\hspace{0.02cm}\includegraphics[scale=0.45]{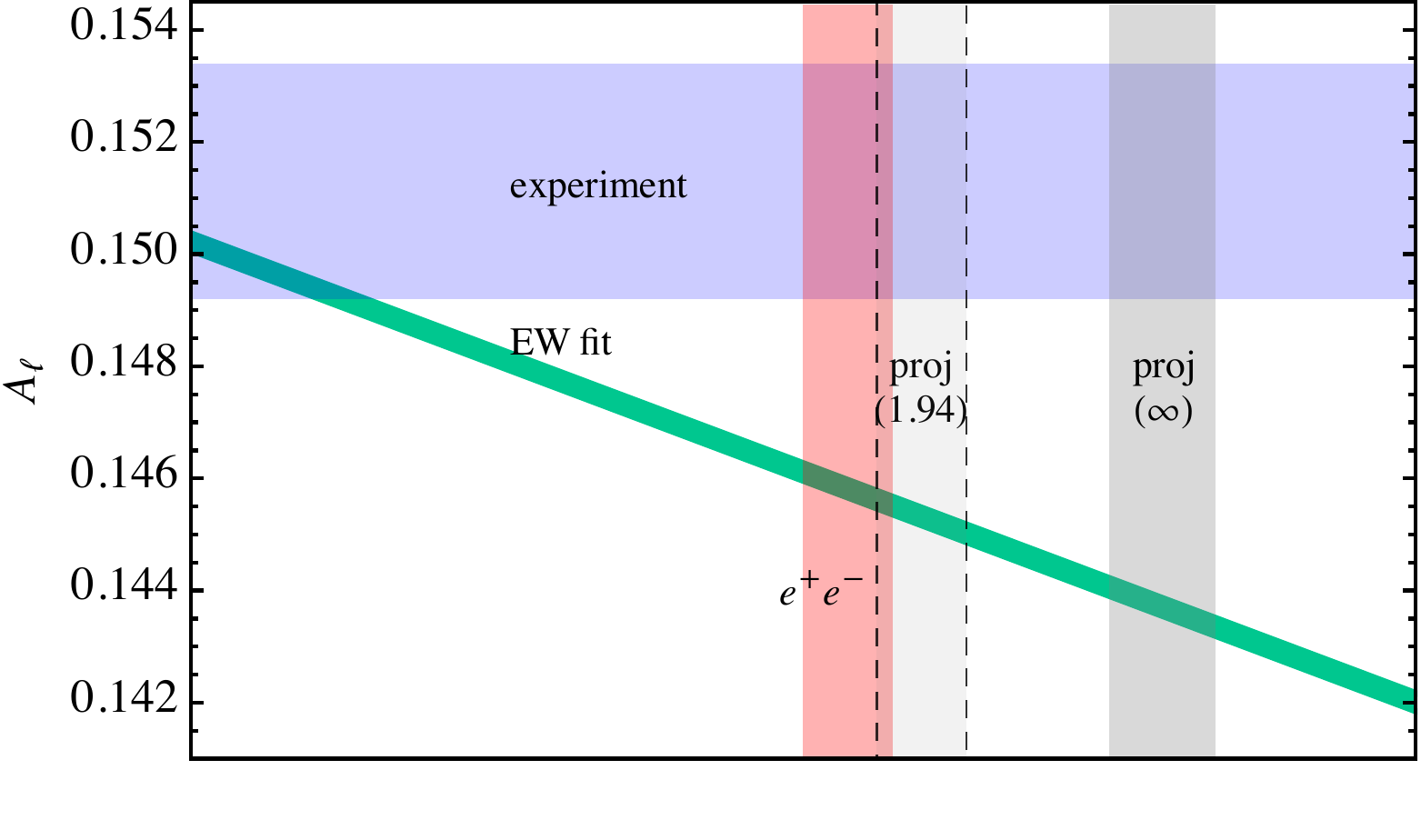}\\[-0.4cm]
\hspace{0.035cm}\includegraphics[scale=0.471]{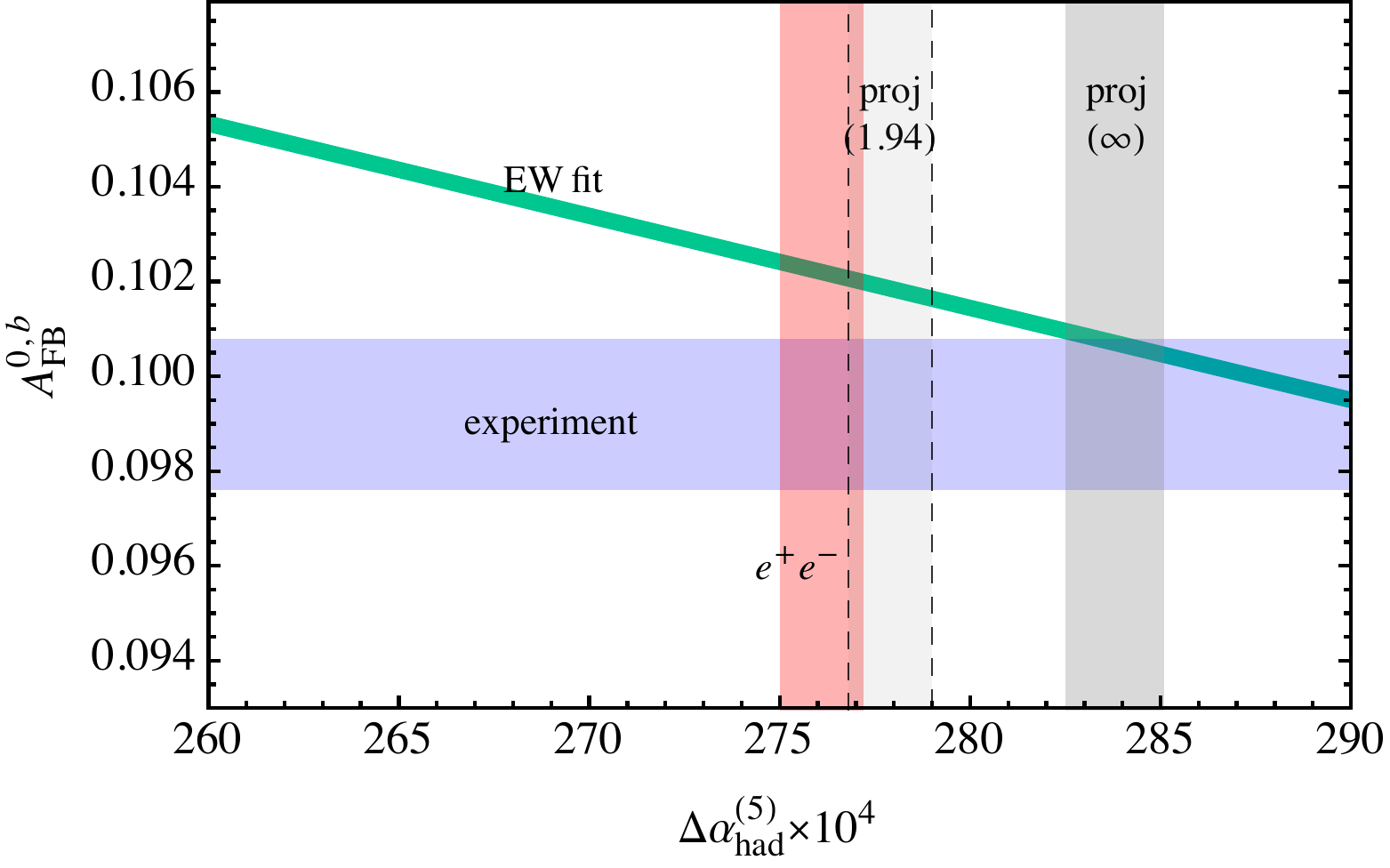}
	\caption{Predictions from the EW fit and measurements for $M_W$, $A_\ell$, and $A_{\rm FB}^{0,b}$ ($1\,\sigma$) as a function of $\Delta \alpha^{(5)}_\text{had}$  together with its preferred ranges from $e^+e^-$ data and the projections~\eqref{BMWc} and~\eqref{BMWclow}. See main text for details.}
	\label{predictions}
\end{figure}

For this purpose, we remove the measurements of three observables with large pulls ($M_W$, $A_\ell$, and $A_{\rm FB}^{0,b}$) from the fit and predict their posterior as a function of $\Delta \alpha^{(5)}_\text{had}$ (without assigning an error to $\Delta \alpha^{(5)}_\text{had}$ for each point sampled). We choose $M_W$ and $A_\ell$ as representatives here given that these are two of the observables that mainly drive the tensions in scenario (3), while the slight improvement in $A_{\rm FB}^{0,b}$ is by far not sufficient to balance their effect. We also note that $A_\ell$ exhibits the biggest tension already in the standard scenario (1), a tension that is further exacerbated in scenario (3). 
The corresponding results are depicted in Fig.~\ref{predictions}, where also the currently preferred ranges for $\Delta \alpha^{(5)}_\text{had}$ as well as the measurements for $M_W$ and $A_\ell$ are included. Therefore, the differences between the posteriors and the measurements, for a given value of $\Delta \alpha^{(5)}_\text{had}$, would need to be explained by BSM physics to restore the goodness of the global EW fit. 
Again, we see that HVP derived from $e^+e^-$ data does not require a BSM component, while for the most extreme projection~\eqref{BMWc} the EW fit is no longer consistent without a significant BSM contribution.

{\it Conclusions and outlook}.---In this Letter we reexamined the impact of HVP on $(g-2)_\mu$ and the global EW fit in light of modern EW precision data. On the one hand, the commonly used result for HVP from $e^+e^-$ data leads to a consistent global EW fit, but generates the well-known discrepancy with the measurement of $(g-2)_\mu$. On the other hand, modifying HVP to render the SM prediction for $(g-2)_\mu$ consistent with the Brookhaven measurement, would not only be in tension with the $e^+e^-$ data, but also increase the tensions within the EW fit, via the change in the hadronic running of the fine-structure constant $\Delta \alpha^{(5)}_\text{had}$. The significance depends on the energy scale where the changes in the cross section occur. Our analysis assumes a naive scaling with respect to the $e^+e^-$ data below different thresholds, see Eqs.~\eqref{BMWc}--\eqref{BMWclow}, representing three qualitatively different ranges in HVP compilations. A similar change of $a_\mu^\text{HVP}$ as studied here was recently suggested by a calculation in lattice QCD~\cite{Borsanyi:2020mff}. If confirmed, our projections for the impact on $\Delta \alpha^{(5)}_\text{had}$ and preliminary results presented in Ref.~\cite{Laurent} suggest that the changes in the cross section would need to be concentrated at very low energies, requiring a large effect in the $2\pi$ channel.\footnote{In fact, the first two space-like bins from Ref.~\cite{Laurent} combined with $\Delta \alpha^{(5)}_\text{had}(M_Z^2) -\Delta \alpha^{(5)}_\text{had}(-M_Z^2)\sim 0.4\times 10^{-4}$~\cite{Keshavarzi:2018mgv} would indicate a scenario close to Eq.~\eqref{BMWclow}.} However, we stress that our results are relevant for any of the forthcoming precision calculations of HVP in lattice QCD, especially in view of the fact that the current average $a_\mu^\text{HVP}=711.6(18.4)\times 10^{-10}$~\cite{Aoyama:2020ynm,Chakraborty:2017tqp,Borsanyi:2017zdw,Blum:2018mom,Giusti:2019xct,Shintani:2019wai,Davies:2019efs,Gerardin:2019rua,Aubin:2019usy,Giusti:2019hkz} also suggests a bigger central value, albeit with sufficiently large uncertainties to be consistent with the $e^+e^-$ value.  

Either way, a significant shift in HVP can in principle account for the experimental value of $(g-2)_\mu$, but at the expense of exacerbating tensions within the EW fit. As seen from Fig.~\ref{predictions}, we observe that for any of the values of $\Delta \alpha^{(5)}_\text{had}$ assumed in Eqs.~\eqref{BMWc}--\eqref{BMWclow}, the shifts predicted by the EW fit for $M_W$ and $A_\ell$ always occur into the direction in which the tension with respect to their measured value increases, an effect much larger than the few shifts in the opposite direction such as for $A_{\rm FB}^{0,b}$. 
These tensions, which, in principle, could end up anywhere between the red and gray bands, would call for an explanation in terms of BSM physics just as the one in $(g-2)_\mu$ would. However, the kind of BSM scenarios required here would be notably different from the ones necessary to explain $(g-2)_\mu$. E.g., a tension in the prediction for $M_W$ with respect to the measured value could be explained in models that generate a sizable effect in the $T$ parameter. Here, composite models (or in the dual picture models with extra dimensions) come to mind. On the other hand, the tension in $g_A^\ell$ could be resolved in models with vector-like leptons. Furthermore, since extra-dimensional or composite models not only lead to sizable effects in the $S$ and $T$ parameters, but also possess vector-like fermions, these models are prime candidates for reconciling the EW fit. However, such a scenario would either imply severe deficiencies in $e^+e^-$ cross sections affecting in the same way different channels measured at different experiments and facilities over decades or some subtle BSM effect in the $e^+e^-$ data. Our analysis thus reaffirms that even if the need for BSM physics were eliminated in $(g-2)_\mu$ by changing HVP, it is likely that other tensions in the SM would arise elsewhere: in the EW fit or, especially if the impact on $\Delta \alpha^{(5)}_\text{had}$ were minimized by concentrating the changes at low energies, in low-energy $e^+e^-$ cross sections.

\begin{acknowledgments}
We thank E.~Bagnaschi for useful discussions concerning the measurement of the $W$ mass and its prediction within the SM and MSSM, Z.~Fodor, L.~Lellouch, and K.~K.~Szab\'o for communication regarding Refs.~\cite{Borsanyi:2017zdw,Borsanyi:2020mff}, and A.~M.~Coutinho for useful discussions regarding \texttt{HEPfit}. Support by the Swiss National Science Foundation, under Project Nos.\ PP00P21\_76884 (A.C., C.M., and M.M.) and PCEFP2\_181117 (M.H.) is gratefully acknowledged.
\end{acknowledgments}

\end{document}